\documentclass[english,aps,reprint,twocolumn,superscriptaddress,longbibliography,showpacs,showkeys]{revtex4-2}
\usepackage[english]{babel}

\usepackage{amsmath,amssymb}
\usepackage{graphicx,xcolor}
\usepackage[unicode,bookmarksnumbered,breaklinks,colorlinks,linktocpage,
citecolor=blue,linkcolor=darkred,urlcolor=darkblue]{hyperref}
\usepackage{bm}
\usepackage{hyperref}
\usepackage{makecell}

\definecolor{darkblue}{rgb}{0.0, 0.0, 0.55}
\definecolor{darkgreen}{rgb}{0.0, 0.2, 0.13}
\definecolor{darkred}{rgb}{0.55, 0.0, 0.0}

\begin{document}

\title{Thermodynamics of the hyperkagome-lattice $S{=}\frac{1}{2}$ Heisenberg ferromagnet}

\author{Maksym Parymuda} 
\email{mparymuda@icmp.lviv.ua} 
\affiliation{Institute for Condensed Matter Physics,
    National Academy of Sciences of Ukraine,
    Svientsitskii Street 1, 
    79011 L'viv, Ukraine}

\author{Taras Krokhmalskii}
\email{krokhm@icmp.lviv.ua}
\affiliation{Institute for Condensed Matter Physics,
	National Academy of Sciences of Ukraine,
	Svientsitskii Street 1, 
	79011 L'viv, Ukraine}

\author{Oleg Derzhko}
\email{derzhko@icmp.lviv.ua}
\affiliation{Institute for Condensed Matter Physics,
	National Academy of Sciences of Ukraine,
	Svientsitskii Street 1, 
	79011 L'viv, Ukraine}
\affiliation{Professor Ivan Vakarchuk Department for Theoretical Physics, 
	Ivan Franko National University of L’viv, 
    Drahomanov Street 12, 
	79005 L’viv, Ukraine}

\date{\today}

\begin{abstract}
The hyperkagome-lattice $S=1/2$ Heisenberg ferromagnet is studied by means of the linear spin-wave theory, the double-time temperature Green's function method,  high-temperature expansions series analysis, and quantum Monte Carlo simulations to examine the effect of lattice geometry on the finite-temperature properties.
In particular, we have found that the Curie temperature $T_c$ for the hyperkagome-lattice (frustrated lattice) ferromagnet is about $0.33\vert J\vert$ that is smaller than $T_c$ for the diamond-lattice (another three-dimensional lattice with the same coordination number 4 but bipartite one) ferromagnet by about 25\%.
\end{abstract}


 
\maketitle

\section{Introduction}
\label{s1}

Frustrated quantum spin systems are a subject of intense ongoing research in the field of magnetism \cite{highmagneticfields2002,quantummagnetism2004,frustratedspinsystems2005,Lacroix2011}. 
Geometric frustration and quantum fluctuations may prevent ground-state ordering even in three dimensions. 
Among several famous examples one could refer to the $S=1/2$ Heisenberg antiferromagnet on the kagome lattice or the pyrochlore lattice in two or three dimensions, respectively. Another interesting three-dimensional lattice is the hyperkagome lattice which has been in focus of several recent studies initiated by experiments on the spinel oxide Na$_4$Ir$_3$O$_8$ \cite{Okamoto2007}.

The present study concerns the $S=1/2$ Heisenberg model on the hyperkagome lattice. However, we focus on the {\em ferromagnetic} sign of the exchange interaction rather than on the antiferromagnetic one.
It is easily seen how the set of Hamiltonian eigenstates depends on the sign of the exchange interaction: Being arranged according to their energy, these states only invert the order under the change of the exchange interaction sign.
As a result, the complicated low-energy states for the antiferromagnet become high-energy states for the ferromagnet and must show up in the finite-temperature properties for the ferromagnet. Previously this has been illustrated for the pyrochlore lattice \cite{mueller2017,hutak2018}.
Now, we are examining the hyperkagome-lattice case.

There are a plenty of methods to investigate the properties of quantum Heisenberg ferromagnets.
Since the ground state is ``all spin up'', a spin-wave theory can be elaborated straightforwardly for examining the low-temperature properties.
Besides, the double-time temperature Green's function method complemented by a rotational-symmetry breaking approximation (like the Tyablikov approximation)
is applicable for ferromagnets.
In addition, 
the high-temperature expansion series for the hyperkagome-lattice $S=1/2$ Heisenberg model are available up to 16th order \cite{Singh2012} (see also Ref.~\cite{Pierre2024}).
Hence, the standard series analysis can be applied to get thermodynamic characteristics \cite{Oitmaa2006,Gonzalez2023}. 
Finally, 
the quantum Monte Carlo method does not suffer from the infamous sign problem for ferromagnets and, for example, the ALPS package \cite{ALBUQUERQUE20071187,Bauer2011} can be utilized for numerical study of the finite-size system thermodynamics. 
By comparing the outcomes of various approximate techniques we can pin down what is really inherent in the model under consideration.
To some extent similar program was realized in Refs.~\cite{hutak2018,Pavizhakumari2025}.

To illustrate effects of the lattice geometry, it is instructive to compare the model in question  to another three-dimensional model with the same coordination number 4 -- the diamond-lattice $S=1/2$ Heisenberg ferromagnet. The latter model has been discussed recently using high-temperature expansion \cite{Oitmaa2018,Kuzmin2019} and quantum Monte Carlo \cite{Baerwolf2025} approaches.  

The remainder of this paper is organized as follows. 
In Sections~\ref{s2} and \ref{s3} we introduce the model and explain the methods to be used.
Then, in Section~\ref{s4}, we report and discuss our main findings.
Finally, we summarize in Section~\ref{s5}.

\section{Model}
\label{s2}

\subsection{Lattice}
\label{s21}

\begin{figure}
\includegraphics[width=0.995\columnwidth]{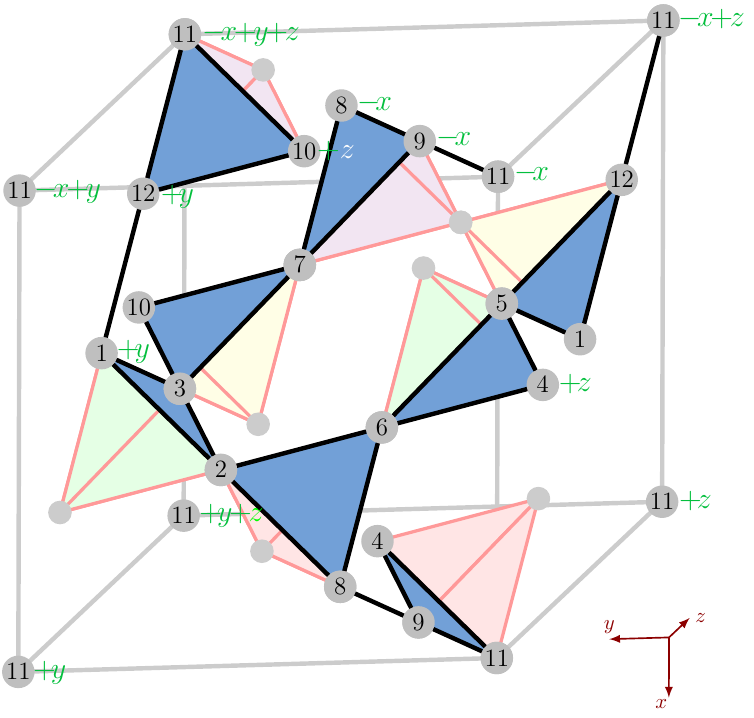} 
\caption{The hyperkagome lattice \cite{Hutak2024}. It has 12 sites in a cubic unit cell (the length of the cube edge is $1$); they are denoted as $1,\ldots,12$. The sites from the neighboring cells acquire, e.g., $-x$ if that cell is shifted by $-1$ in $x$ direction, $+y$ if that cell is shifted by $+1$ in $y$ direction, and so on. For further explanation see the main text.}
\label{fig01}
\end{figure}

The hyperkagome lattice is a three-dimensional network of corner-sharing triangles with 12 sites in the cubic unit cell, see Fig.~\ref{fig01}. 
More precisely, 
the sites on the hyperkagome lattice may be defined by 
$\bm{R}_{\bm{m}\alpha}=\bm{R}_{\bm{m}}+\bm{r}_{\alpha}$.
Here
$\bm{R}_{\bm{m}}=m_{x}\bm{e}_{x}+m_{y}\bm{e}_{y}+m_{z}\bm{e}_{z}$,
where $m_{x},m_{y},m_{z}$ are integers 
and 
$\bm{e}_{x}=(1,0,0)$, $\bm{e}_{y}=(0,1,0)$, $\bm{e}_{z}=(0,0,1)$. 
Furthermore,
the origins of twelve equivalent sublattices are defined by 
$\bm{r}_{\alpha}$, $\alpha=1,2,\dots,12$, 
where
\begin{eqnarray}
\label{01}
\bm{r}_{1}{=}\frac{1}{4}(-2, 0, 2),
\;
\bm{r}_{2}{=}\frac{1}{4}(-1, 3, 2),
\;
\bm{r}_{3}{=}\frac{1}{4}(-2, 3, 1),
\nonumber\\
\bm{r}_{4}{=}\frac{1}{4}(-1, 1, 0),
\;
\bm{r}_{5}{=}\frac{1}{4}(-2, 1, 3),
\;
\bm{r}_{6}{=}\frac{1}{4}(-1, 2, 3),
\nonumber\\
\bm{r}_{7}{=}\frac{1}{4}(-3, 2, 1),
\;
\bm{r}_{8}{=}\frac{1}{4}( 0, 2, 2),
\;
\bm{r}_{9}{=}\frac{1}{4}( 0, 1, 1),
\nonumber\\
\bm{r}_{10}{=}\frac{1}{4}(-3, 3, 0),
\;
\bm{r}_{11}{=}( 0, 0, 0),
\;
\bm{r}_{12}{=}\frac{1}{4}(-3, 0, 3).	
\end{eqnarray}
In Fig.~\ref{fig01}, the site $n_x{=}n_y{=}n_z{=}0$, $\bm{r}_{1}$ is denoted as 1, the site $n_x{=}-1$, $n_y{=}n_z{=}1$, $\bm{r}_{11}$, i.e., ${-}\bm{e}_{x}{+}\bm{e}_{y}{+}\bm{e}_{z}{+}\bm{r}_{11}$, is denoted as $11{-}x{+}y{+}z$, and so on.
Each site has 4 nearest neighbors; the neighboring sites are connected by a bond. Therefore, we have to consider 24 bonds:
15 bonds connecting the sites inside the unit cell,
$1-5$, $1-12$, $2-3$, $2-6$, $2-8$, $3-7$, $3-10$, $4-9$, $4-11$, $5-6$, $5-12$, $6-8$, $7-10$, $8-9$, $9-11$,
and
9 bonds connecting the sites from the unit cell $n_x{=}n_y{=}n_z{=}0$ to the sites from the neighboring unit cells, 
$1-(2{-}y)$, $1-(3{-}y)$, $4-(5{-}z)$, $4-(6{-}z)$, $7-(8{-}x)$, $7-(9{-}x)$, $10-(11{-}x{+}y)$, $10-(12{+}y{-}z)$, $11-(12{+}x{-}z)$.
In Fig.~\ref{fig01} we show for clarity 28 bonds (thick black), i.e., besides the 15 bonds inside the unit cell, the following bonds: 
$(1{+}y)-2$ (equivalent to $1-(2{-}y)$), 
$(1{+}y)-3$ (equivalent to $1-(3{-}y)$), 
$(1{+}y)-(12{+}y)$ 
(cf. $1-12$), 
$(4{+}z)-5$ (equivalent to $4-(5{-}z)$), 
$(4{+}z)-6$ (equivalent to $4-(6{-}z)$), 
$7-(8{-}x)$, 
$7-(9{-}x)$, 
$(8{-}x)-(9{-}x)$ 
(cf. $8-9$), 
$(9{-}x)-(11{-}x)$ 
(cf. $9-11$),
$(10{+}z)-(11{-}x{+}y{+}z)$ (equivalent to $10-(11{-}x{+}y)$), 
$(10{+}z)-(12{+}y)$ (equivalent to $10-(12{+}y{-}z)$), 
$(11{-}x{+}z)-12$ (equivalent to $11-(12{+}x{-}z)$), 
$(11{-}x{+}y{+}z)-12{+}y$ 
(cf. $11-(12{+}x{-}z)$ or $(11{-}x{+}z)-12$).

In Fig.~\ref{fig01} we also show the underlying pyrochlore lattice which can be used to obtain the hyperkagome lattice after removing spins from one out of the four sites in every tetrahedron.

\subsection{Spin Hamiltonian}
\label{s22}

In the present study,
we consider the $S=1/2$ ferromagnetic Heisenberg model on the hyperkagome lattice 
given by the Hamiltonian
\begin{equation}
\label{02}
H=-\vert J\vert\sum_{\langle\bm{m}\alpha;\bm{n}\beta\rangle}
\bm{S}_{\bm{m}\alpha}\cdot\bm{S}_{\bm{n}\beta}.
\end{equation}
We set the ferromagnetic interaction constant $\vert J\vert=1$ this way fixing the energy scale. 
The sum in Eq.~(\ref{02}) runs over all edges of the hyperkagome lattice; these bonds, which join the neighboring sites, are shown in Fig.~\ref{fig01} by black segments. 
The Heisenberg coupling can be also written as $\bm{S}_{\bm{m}\alpha}\cdot\bm{S}_{\bm{n}\beta}=({S}^-_{\bm{m}\alpha}{S}^+_{\bm{n}\beta}+{S}^+_{\bm{m}\alpha}{S}^-_{\bm{n}\beta})/2+{S}^z_{\bm{m}\alpha}{S}^z_{\bm{n}\beta}$ and ${S}^z=1/2-S^-S^+$.

For analytical calculations it is convenient to impose periodic boundary conditions  and introduce the following operators
\begin{eqnarray}
\label{03}
S^+_{{\bm q}\alpha}
{=}\frac{1}{\sqrt{{\cal N}}}\sum_{{\bm m}}{\rm e}^{{-}{\rm i}{\bm q}{\cdot}{\bm R}_{\bm m}}S^+_{{\bm m}\alpha},
\nonumber\\
S^-_{{\bm q}\alpha}
{=}\frac{1}{\sqrt{{\cal N}}}\sum_{{\bm m}}{\rm e}^{{\rm i}{\bm q}{\cdot}{\bm R}_{\bm m}}S^-_{{\bm m}\alpha}.
\end{eqnarray}
Here
$\alpha=1,\ldots,12$,
${\cal N}=N/12$ is the number of unit cells,
${\cal N}={\cal L}_x{\cal L}_y{\cal L}_z$,
${\bm q}{\cdot}{\bm R}_{\bm m}=q_xm_x+q_ym_y+q_zm_z$,
where
$q_x=2\pi z_x/{\cal{L}}_x$, $z_x=1,\ldots,{\cal {L}}_x$, and so on.
Periodic boundary conditions are used in computer simulations, too.

\section{Methods}
\label{s3}

\subsection{Linear spin-wave theory}
\label{s31}

\begin{figure}
\includegraphics[width=0.995\columnwidth]{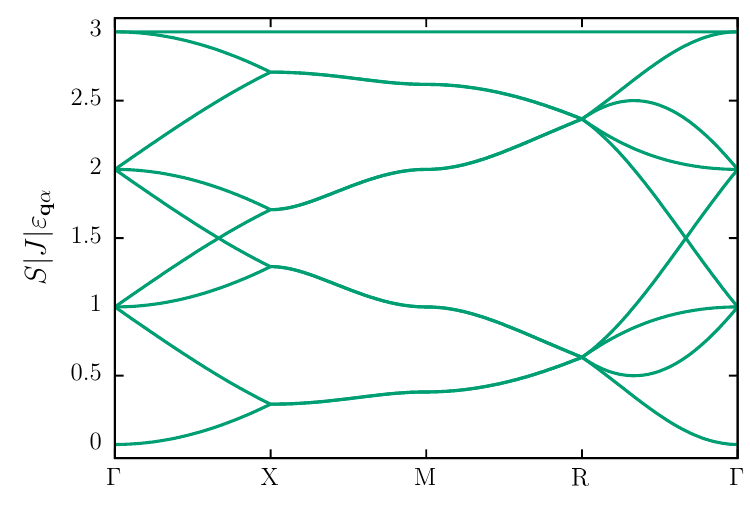} 
\caption{One-magnon bands $S\vert J\vert\varepsilon_{{\bm q}\alpha}$ along the path $\Gamma - X - M - R - \Gamma$ for the hyperkagome-lattice $S=1/2$ Heisenberg ferromagnet. 
The special points in the first Brillouin zone of a simple cubic lattice are defined as follows:
$\Gamma=(0, 0, 0)$, $X=(0, \pi, 0)$, $M=(\pi,\pi,0)$, and $R=(\pi,\pi,\pi)$.
There are 12 bands in total: 
8 bands are dispersive with the energy $0\le S\vert J\vert\varepsilon_{{\bm q}\alpha}\le 3$, $\alpha=1,\ldots,8$
and
4 bands are dispersionless (flat) with the energy $S\vert J\vert\varepsilon_{{\bm q}\alpha}=3$, $\alpha=9,\ldots,12$.}
\label{fig02}
\end{figure}

A simplest approach to obtain the low-temperature thermodynamics of a Heisenberg ferromagnet is the linear spin-wave theory. Using the Holstein-Primakoff transformation \cite{Holstein1940}
\begin{eqnarray}
\label{04}
S^+\approx \sqrt{2S}a,
\;\;\;
S^-\approx \sqrt{2S}a^\dagger,
\;\;\;
S^z=S-a^\dagger a
\end{eqnarray}
(in the end, we set $S=1/2$)
we get for each bond
\begin{eqnarray}
\label{05}
\bm{S}_{\bm{m}\alpha}\cdot\bm{S}_{\bm{n}\beta}
\nonumber\\
{\approx}
S^2{-}S\left(
a^\dagger_{\bm{m}\alpha} a_{\bm{n}\beta}
{+}
a_{\bm{m}\alpha} a^\dagger_{\bm{n}\beta} 
{-} a^\dagger_{\bm{m}\alpha}a_{\bm{m}\alpha}{-} a^\dagger_{\bm{n}\beta}a_{\bm{n}\beta}\right)
\nonumber\\
{+}a^\dagger_{\bm{m}\alpha}a_{\bm{m}\alpha}a^\dagger_{\bm{n}\beta}a_{\bm{n}\beta}.
\end{eqnarray}
Note that the last term in Eq.~(\ref{05}) may be omitted as irrelevant for determination of the one-magnon spectrum.
Next, for the model under consideration, we introduce 12 bosonic operators
\begin{eqnarray}
\label{06}
a_{{\bm q}\alpha}
=\frac{1}{\sqrt{{\cal N}}}\sum_{{\bm m}}{\rm e}^{-{\rm i}{\bm q}\cdot{\bm R}_{\bm m}}a_{{\bm m}\alpha},
\;\;\;
\alpha=1,\ldots,12,
\end{eqnarray}
cf. Eq.~(\ref{03}).
Furthermore,
for 15 bonds connecting the sites within the same unit cell we have
\begin{eqnarray}
\label{07}
\sum_{\bm m} \bm{S}_{\bm{m}\alpha}\cdot\bm{S}_{\bm{m}\beta}
\rightarrow
{\cal N}S^2
\nonumber\\
{+}
S\sum_{\bm q}
\left(
a_{\bm{q}\alpha}^\dagger a_{\bm{q}\beta}
{+} 
a_{\bm{q}\alpha} a^\dagger_{\bm{q}\beta}
{-}a_{\bm{q}\alpha}^\dagger a_{\bm{q}\alpha}{-}a_{\bm{q}\beta}^\dagger a_{\bm{q}\beta}
\right).
\end{eqnarray}
Whereas for 9 bonds connecting the sites from neighboring unit cells,
i.e., 
$1-(2{-}y)$, 
$1-(3{-}y)$, 
$4-(5{-}z)$, 
$4-(6{-}z)$, 
$7-(8{-}x)$, 
$7-(9{-}x)$, 
$10-(11{-}x{+}y)$, 
$10-(12{+}y{-}z)$, 
$11-(12{+}x{-}z)$
we have,
for instance,
\label{08}
\begin{eqnarray}
\sum_{\bm m} \bm{S}_{\bm{m}1}\cdot\bm{S}_{\bm{m}-\bm{e}_y,2}
\rightarrow
{\cal N}S^2
\nonumber\\
{+}
S\sum_{\bm q}
\left(
a_{\bm{q}1}^\dagger a_{\bm{q}2}{\rm e}^{-{\rm i}q_y}
{+} 
a_{\bm{q}1} a^\dagger_{\bm{q}2}{\rm e}^{{\rm i}q_y}
{-}a_{\bm{q}1}^\dagger a_{\bm{q}1}{-}a_{\bm{q}2}^\dagger a_{\bm{q}2}
\right),
\end{eqnarray}
and so on.
As a result, the Hamiltonian of the spin model becomes
\begin{eqnarray}
\label{09}
H\rightarrow
-24{\cal N}S^2 \vert J\vert 
\nonumber\\
+S\vert J\vert
\sum_{{\bm q}}
\left(
\begin{array}{ccc}
a_{{\bm q}1}^\dagger & \ldots & a_{{\bm q}12}^\dagger
\end{array}
\right)
{\bm F}
\left(
\begin{array}{c}
a_{{\bm q}1} \\
\vdots \\
a_{{\bm q}12}
\end{array}
\right),
\end{eqnarray} 
where 
\begin{widetext}
\begin{eqnarray}
\label{10}
F_{\alpha\alpha}=4,
\nonumber\\
F_{12}=F_{13}=-{\rm e}^{-{\rm i}q_y}, 
\;
F_{15}=F_{1,12}=-1,
\nonumber\\
F_{23}=F_{26}=F_{28}=F_{37}=F_{3,10}=-1,
\nonumber\\
F_{45}=F_{46}=-{\rm e}^{-{\rm i}q_z},
\;
F_{49}=F_{4,11}=-1,
\nonumber\\
F_{5,12}=F_{56}=F_{68}=-1,
\nonumber\\
F_{78}=F_{79}=-{\rm e}^{-{\rm i}q_x},
\;
F_{7,10}=F_{89}=F_{9,11}=-1,
\nonumber\\
F_{10,11}{=}{-}{\rm e}^{{\rm i}(q_y{-}q_x)},
F_{10,12}{=}{-}{\rm e}^{{\rm i}(q_y{-}q_z)},
F_{10,11}{=}{-}{\rm e}^{{\rm i}(q_x{-}q_z)};
\end{eqnarray}
\end{widetext}
other matrix elements are zero
and
$F_{\alpha\beta}=F_{\beta\alpha}^*$.
The Hermitian matrix ${\bm F}$ can be brought into the diagonal form by a unitary transformation,
\begin{eqnarray}
\label{11}
{\bm U}{\bm F}{\bm U}^\dagger=
\left(
\begin{array}{ccc}
\varepsilon_{{\bm q}1} & \ldots & 0\\
\vdots & \vdots & \vdots \\
0 &\ldots  & \varepsilon_{{\bm q}12}
\end{array}
\right).
\end{eqnarray}
The one-magnon energies $S\vert J\vert\varepsilon_{{\bm q}\alpha}$, $\alpha=1,\ldots,12$ are shown in Fig.~\ref{fig02}.  

Thermodynamic properties of noninteracting bosons with the Hamiltonian
\begin{eqnarray}
\label{12}
H=-24{\cal N}S^2\vert J\vert + S\vert J\vert\sum_{{\bm q}\alpha} \varepsilon_{{\bm q}\alpha} \eta_{{\bm q}\alpha}^\dagger \eta_{{\bm q}\alpha}
\end{eqnarray}
can be easily calculated; 
they yield the low-temperature properties of the quantum spin system (\ref{02}). 
Thus ($k_{\rm B}=1$),
the internal energy $E=Ne$ is given by
\begin{eqnarray}
\label{13}
E=-24{\cal N}S^2\vert J\vert +S\vert J\vert\sum_{{\bm q} \alpha}\frac{\varepsilon_{{\bm q}\alpha}}{{\rm e}^{\frac{S\vert J\vert\varepsilon_{{\bm q}\alpha}}{T}}-1},
\end{eqnarray}
whereas the magnetization $M=Nm$ is given by
\begin{eqnarray}
\label{14}
M=12{\cal N}S-\sum_{{\bm q} \alpha}\frac{1}{{\rm e}^{\frac{S\vert J\vert\varepsilon_{{\bm q}\alpha}}{T}}-1}.
\end{eqnarray}
As a result, the specific heat is given by $C=\partial E/\partial T$ and the entropy $S=\int_0^T{\rm d}T C/T$. 

All above expressions can be evaluated numerically. Here, we have to use in Eqs.~(\ref{13}), (\ref{14}) that in the thermodynamic limit $\sum_{{\bm q}}(\ldots)/{\cal N}$ becomes $\int_{-\pi}^{\pi}{\rm d}q_x\int_{-\pi}^{\pi}{\rm d}q_y\int_{-\pi}^{\pi}{\rm d}q_z(\ldots)/(8\pi^3)$.
Clearly, since for the lowest-energy (acoustic) magnon band $\varepsilon_{{\bm q}\to 0}\to {\bm q}^2/16$ (see also Fig.~\ref{fig02}), all thermodynamic quantities show the standard behavior as $T\to 0$, i.e., $E\propto T^{5/2}$, $C\propto T^{3/2}$, $NS-M\propto T^{3/2}$. 
In Sec.~\ref{s4}, we report $C/N$ and $M/N$ for $T\le 0.2$.

\subsection{Green's function method: Mean-field and random-phase approximations}
\label{s32}

Now, we turn to the double-time temperature Green's function method.
We introduce the (retarded) Green's functions ($\hbar=1$) \cite{Zubarev1960}
\begin{eqnarray}
\label{15}
G_{\alpha\beta}(t)\equiv\langle\langle S^+_{{\bm q}\alpha}\vert S^-_{{\bm q}\beta}\rangle\rangle_t
=-{\rm i}\Theta(t)\langle [S^+_{{\bm q}\alpha}(t), S^-_{{\bm q}\beta}]\rangle,	
\nonumber\\
G_{\alpha\beta}
\equiv
G_{\alpha\beta}(\omega)=\int\limits_{-\infty}^{\infty}{\rm d}t{\rm e}^{{\rm i}\omega t}G_{\alpha\beta}(t),
\end{eqnarray} 
where the operators $S_{{\bm q}\alpha}^{\pm}$ are defined in Eq.~(\ref{03}).
$G_{\alpha\beta}(\omega)$ gives immediately the dynamic susceptibility $\chi_{\bm q}^{+-}(\omega)=-\sum_{\alpha,\beta=1}^{12}G_{\alpha\beta}(\omega)/12$ and the correlation functions $\langle S_{{\bm q}\beta}^{-}(t)S_{{\bm q}\alpha}^{+}\rangle=[{\rm i}/(2\pi)]\lim_{\epsilon\to 0}\int_{-\infty}^{\infty}{\rm d}\omega{\rm e}^{-{\rm i}\omega t}[G_{\alpha\beta}(\omega+{\rm i}\epsilon)-G_{\alpha\beta}(\omega-{\rm i}\epsilon)]/({\rm e}^{\omega/T}-1)$. Equal-time correlations yield thermodynamics. In particular, the magnetization per site $m=\langle S^z\rangle$ is given by $\langle S^z\rangle=1/2-(1/N)\sum_{{\bm q}\alpha}\langle S_{{\bm q}\alpha}^{-}S_{{\bm q}\alpha}^{+}\rangle$. 
To calculate the Green's function, we use the equation-of-motion method \cite{Zubarev1960}.

To begin, we consider the simplest mean-field version of the Hamiltonian (\ref{02})
\begin{eqnarray}
\label{16}
H=2N\vert J\vert\langle S^z\rangle^2
-4\vert J\vert\langle S^z\rangle\sum_{{\bm m}}\sum_{\alpha}S^z_{{\bm m}\alpha}.
\end{eqnarray}
The exact first-order equation of motion gives 
\begin{eqnarray}
\label{17}
G_{\alpha\beta}
=\frac{2\langle S^z\rangle \delta_{\alpha\beta}}{\omega- 4\vert J\vert \langle S^z\rangle},
\;\;\;
\langle S^-_{{\bm q}\beta}S^+_{{\bm q}\alpha}\rangle=\frac{2\langle S^z\rangle\delta_{\alpha\beta}}{{\rm e}^{\frac{4\vert J\vert\langle S^z\rangle}{T}}-1},
\end{eqnarray}
and, as a result, $\langle S^z\rangle$ satisfies the self-consistent equation
\begin{eqnarray}
\label{18}
1=2\langle S^z\rangle\coth \frac{2\vert J\vert\langle S^z\rangle}{T}.
\end{eqnarray}
This equation yields the mean-field prediction for the Curie temperature $T_c/\vert J\vert=1$, which depends only on the number of the nearest neighbors, with $\langle S^z\rangle\le 1/2$ for $T<T_c$ but $\langle S^z\rangle=0$ for $T\ge T_c$, see Sec.~\ref{s4}. 
Thermodynamics within the mean-field approximation can be obtained from the internal energy $E=-2N\vert J\vert \langle S^z\rangle^2$.

Next, we do not break explicitly the rotational-invariant symmetry of the spin Hamiltonian (\ref{02}) as in Eq.~(\ref{16}), but write down the first-order equation of motion, and calculate the following time derivative
\begin{eqnarray}
\label{19}
\dot{S}^+_{{\bm q}\alpha}
{=}\frac{{\rm i}}{\sqrt{{\cal N}}}\sum_{{\bm m}}{\rm e}^{{-}{\rm i}{\bm q}{\cdot}{\bm R}_{\bm m}}\left[H,S^+_{{\bm m}\alpha}\right].
\end{eqnarray}
Note that each site has only 4 neighboring sites.
Furthermore,
$[{\bm S}_A{\cdot}{\bm S}_B, S_B^+]=-S_A^+S_B^z+S_A^zS_B^+\to \langle S^z\rangle\left(-S_A^++S_B^+\right)$ (Tyablikov approximation).
As a result, we have, e.g.,
\begin{eqnarray}
\label{20}
\omega G_{1\beta}=2\langle S^z\rangle\delta_{1\beta}
\nonumber\\
{+}\vert J\vert\langle S^z\rangle\!\left(4G_{1\beta}
{-}{\rm e}^{{-}{\rm i}q_y}G_{2\beta}{-}{\rm e}^{{-}{\rm i}q_y}G_{3\beta}
{-}G_{5\beta}{-}G_{12,\beta}
\right),
\nonumber\\
\end{eqnarray}
and so on or in the matrix form
\begin{eqnarray}
\label{21}
\left(\omega {\bm 1}-\vert J\vert\langle S^z\rangle{\bm F}\right)\,{\bm G}=2\langle S^z\rangle{\bm 1},
\end{eqnarray}
where the matrix ${\bm F}$ is defined in Eq.~(\ref{10}).
Knowing the eigenvectors $\langle \beta\vert {\bm q}\gamma\rangle$, $\beta=1,\ldots,12$ and the corresponding eigenvalues $\varepsilon_{{\bm q}\gamma}$ of the matrix ${\bm F}$,
i.e.,
$\sum_{\beta}F_{\alpha\beta}\langle\beta\vert{\bm q}\gamma\rangle=\varepsilon_{{\bm q}\gamma}\langle\alpha\vert{\bm q}\gamma\rangle$,
we can get the desired result
\begin{eqnarray}
\label{22}
G_{\alpha\beta}=2\langle S^z\rangle\sum_{\gamma=1}^{12}\frac{\langle\alpha\vert {\bm q}\gamma \rangle\langle {\bm q}\gamma \vert\beta\rangle}{\omega - \vert J\vert \langle S^z\rangle \varepsilon_{{\bm q}\gamma}}.
\end{eqnarray} 
The eigenvectors $\langle \alpha\vert {\bm q}\gamma\rangle$ and $\langle {\bm q}\gamma \vert\beta\rangle$ appearing in Eq.~(\ref{22}) are not explicitly calculated, since $\langle S^z\rangle$ contains after all $\sum_{\alpha}\langle \alpha\vert {\bm q}\gamma\rangle\langle {\bm q}\gamma\vert \alpha\rangle =1$. Therefore, we arrive at the following equation for $\langle S^z\rangle$:
\begin{eqnarray}
\label{23}
\langle S^z\rangle=\frac{1}{2}-\frac{\langle S^z\rangle}{6{\cal{N}}}\sum_{\gamma=1}^{12}\sum_{{\bm q}}
\frac{1}{{\rm e}^{\frac{\vert J\vert\langle S^z\rangle\varepsilon_{{\bm q}\gamma}}{T}}-1}.
\end{eqnarray}
Equation for $\langle S^z\rangle$ (\ref{23}) at $T_c$, where $\langle S^z\rangle$ vanishes, yields the following Curie temperature $T_c$:
\begin{eqnarray}
\label{24}
\frac{T_c}{\vert J\vert}=
\frac{3}{\sum_{\gamma=1}^{12}\frac{1}{{\cal N}}\sum_{{\bm q}}\frac{1}{\varepsilon_{{\bm q}\gamma}}}
\approx 0.433\,8.
\end{eqnarray}
Again, we have to use that in the thermodynamic limit $\sum_{{\bm q}}(\ldots)/{\cal N}$ becomes $\int_{-\pi}^{\pi}{\rm d}q_x\int_{-\pi}^{\pi}{\rm d}q_y\int_{-\pi}^{\pi}{\rm d}q_z(\ldots)/(8\pi^3)$. We illustrate a numerical solution of Eq.~(\ref{23}) in Sec.~\ref{s4}.
Further analysis of thermodynamics within the Tyablikov approximation can be performed as in Ref.~\cite{hutak2018}, where the pyrochlore-lattice case was examined.

\subsection{High-temperature expansion series and the Curie temperature}
\label{s33}

Using the high-temperature expansion series with respect to $\beta=1/T$ reported in Ref.~\cite{Singh2012},
that is,
\begin{eqnarray}
\label{25}
\frac{\chi(\beta)}{\beta}=\sum_{n=0}^{16}\frac{b_n}{4^n}\beta^n
\end{eqnarray}
with the series expansion coefficients $b_n$ given in Table~I of Ref.~\cite{Singh2012},
we can find the Curie temperature $T_c=1/\beta_c$, see, e.g., Refs.~\cite{Oitmaa2006,Gonzalez2023}. 
According to the Dlog Pad\'{e} method \cite{Gonzalez2023}, since $\chi\propto (\beta_c - \beta)^{-\gamma}$ for $\beta \approx \beta_c$, $\beta < \beta_c$ ($\gamma>0$ is a critial exponent), one has
$\chi(\beta)/\chi^{\prime}(\beta)=(\beta_c - \beta)/\gamma$.
Now a ratio of two polynomials $\chi(\beta)$ and $\chi^{\prime}(\beta)$ obtained according to  Eq.~(\ref{25}) is re-expanded in powers of $\beta$. For the resulting polynomial of order 16 we construct a number of Pad\'{e} approximants $[u, d]$ to this result, $u+d \le 16$, and seek for each of them the least positive root, $\beta_c$ and the slope at $\beta_c$.

More specifically \cite{Parymuda2024}, we estimate $T_c=1/\beta_c$ and $\gamma$ from $[u, d]$ Pad\'{e} approximants to the series $\chi(\beta)/\chi^{\prime}(\beta)$ with $d=4,5,6,7,8$ and $u=6,7,8$. The arithmetic mean of 15 values of $T_c$ gives $T_c\approx 0.326$. If we drop out two largest deviations from the mean value and then find the arithmetic mean of 13 values of $T_c$ we get $T_c\approx 0.338\pm 0.044$.
For $\gamma$ we get only a rough estimate, $\gamma\approx 1.44$, since the corresponding entries in the Pad\'{e} table are not very close to each other. 

\subsection{Quantum Monte Carlo simulations}
\label{s34}

\begin{figure}
\includegraphics[width=0.995\columnwidth]{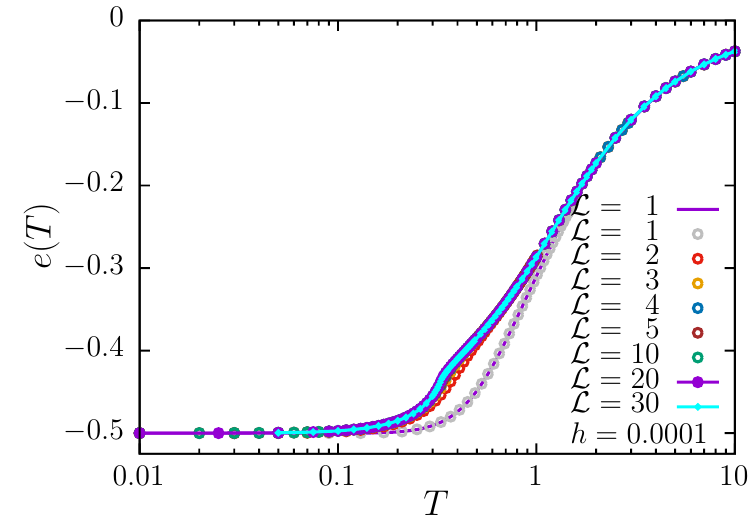} 
\includegraphics[width=0.995\columnwidth]{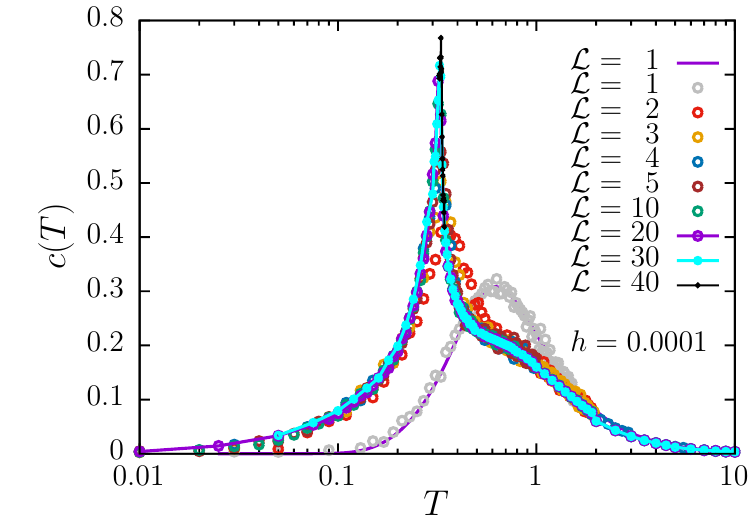} 
\includegraphics[width=0.995\columnwidth]{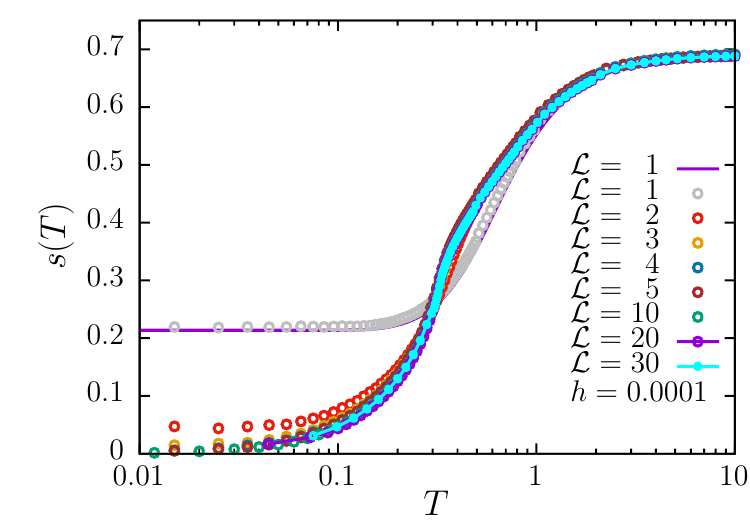} 
\caption{Thermodynamics of the hyperkagome-lattice $S=1/2$ Heisenberg ferromagnet.
Quantum Monte Carlo simulations for (from top to bottom) internal energy, specific heat, and entropy. The value of the symmetry-breaking field is $h=10^{-4}$.}
\label{fig03}
\end{figure}

\begin{figure}
\includegraphics[width=0.995\columnwidth]{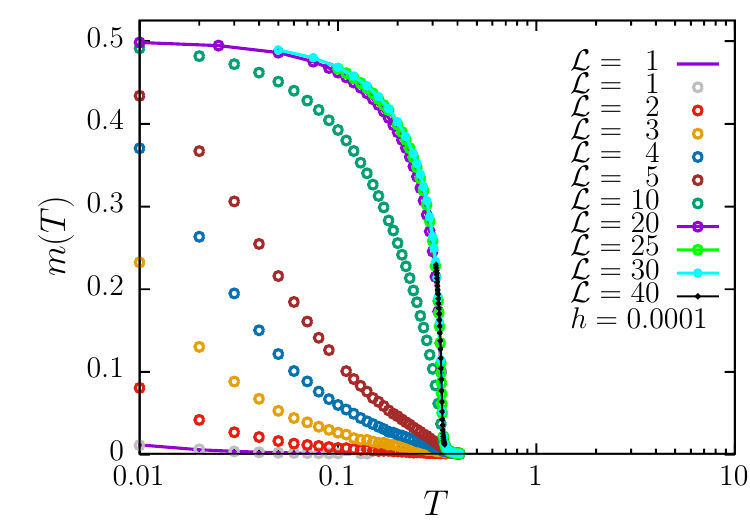} 
\includegraphics[width=0.995\columnwidth]{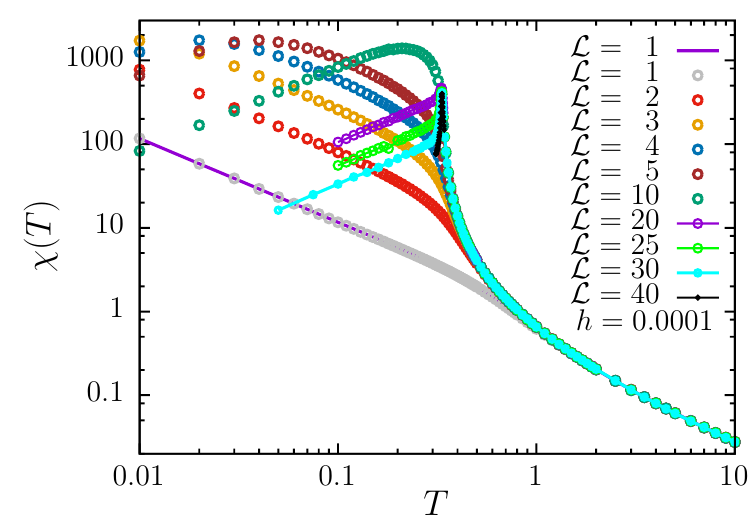}
\includegraphics[width=0.995\columnwidth]{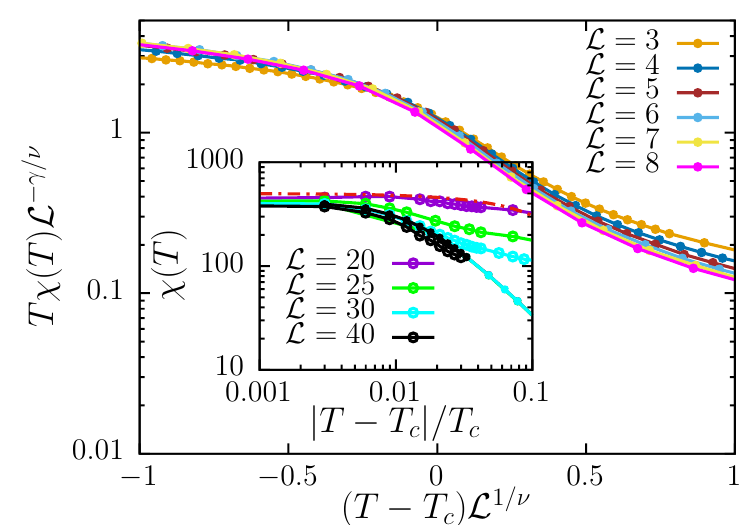} 
\caption{Thermodynamics of the hyperkagome-lattice $S=1/2$ Heisenberg ferromagnet.
Quantum Monte Carlo simulations for (top) magnetization and (middle) susceptibility. In the bottom panel we illustrate the finite-size scaling for $\chi(T)$. Here $\chi(T)$ is calculated for various size ${\cal L}$, $T_{c}=0.334$, $\gamma=1.39$, $\nu=0.71$. In the inset we show how $\chi(T)$ approaches critical behavior when $\vert \tau\vert\to 0$, $\tau =(T-T_c)/T_c$ below (filled circles) and above (empty circles) $T_c$ as ${\cal L}$ grows; the dash-dotted curve denotes $500\vert\tau\vert^{-\gamma}$. The value of the symmetry-breaking field is $h=10^{-4}$.}
\label{fig04}
\end{figure}

\begin{figure}
\includegraphics[width=0.995\columnwidth]{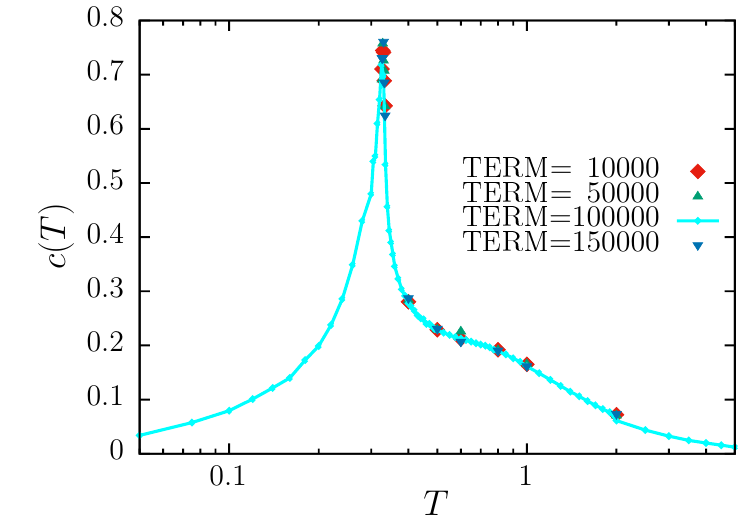} 
\includegraphics[width=0.995\columnwidth]{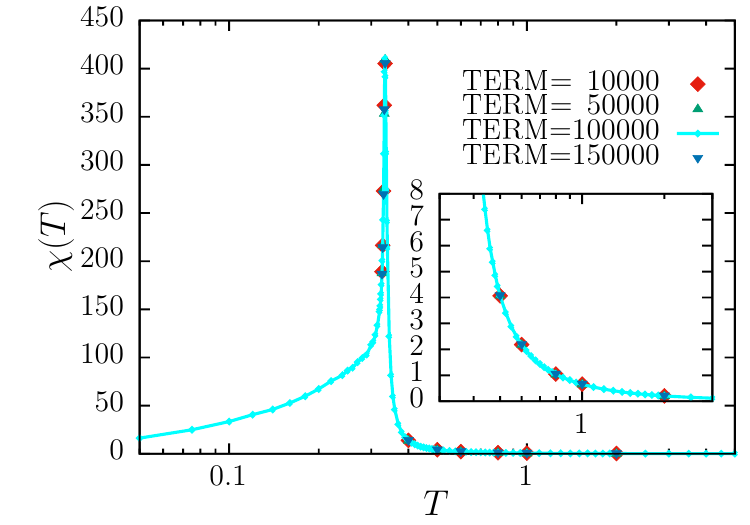}
\caption{Inspecting equilibration at different temperatures $T=0.326$, $0.328$, $0.33$, $0.332$, $0.334$, $0.4$, $0.5$, $0.6$, $0.8$, $1$, $2$, ${\cal L}=30$, $h=10^{-4}$. To realize the equilibrium state, we made $10\,000$, $50\,000$, $100\,000$, $150\,000$ Monte Carlo steps. Then, we took the thermal average to get (top) the specific heat or (bottom) the susceptibility. Inset in the bottom panel is a zoom-in of the temperature region $T=0.3\ldots3$.}
\label{fig05}
\end{figure}

In our study,
we also use the ALPS package (directed-loop scheme in SSE method) \cite{ALBUQUERQUE20071187,Bauer2011} to perform quantum Monte Carlo simulations, see Figs.~\ref{fig03}, \ref{fig04}, and \ref{fig05}. 
We consider periodic lattices of ${\cal N}={\cal L}^3$ unit cells with ${\cal L}=1$, $2$, $3$, $4$, $5$, $6$, $7$, $8$, $10$, $20$, $25$, $30$, and $40$.
In our simulations, we introduce a small (symmetry-breaking) magnetic field along $z$ direction with $h=10^{-4}$, $10^{-5}$, or $10^{-6}$ and examine an interplay of the values of $h$ and ${\cal N}$ on the desired temperature profiles.
From the ALPS-package computations we obtain temperature dependencies of the internal energy $E$, the magnetization $\langle \sum_jS_j^z\rangle$, and $\langle(\sum_jS_j^z)^2\rangle$. Then we calculate the specific heat, the entropy, and the susceptibility:
\begin{eqnarray}
\label{26}
C=\frac{\partial E}{\partial T},
\nonumber\\
S=\int\limits_{0}^{T}{\rm d}T\frac{C(T)}{T}=N\ln 2-\int\limits_{T}^{\infty}{\rm d}T\frac{C(T)}{T},
\nonumber\\
N\chi=\frac{\langle(\sum_jS_j^z)^2\rangle-\langle \sum_jS_j^z\rangle^2}{T}.
\end{eqnarray} 
In the performed quantum Monte Carlo simulations, we usually have 100\,000 of thermalisation steps before computing averages by averaging over next 100\,000 steps. We vary the number of thermalisation steps, $10\,000$, $50\,000$, $100\,000$, $150\,000$, to check equilibration at different temperatures, $T=0.326$, $0.328$, $0.33$, $0.332$, $0.334$, $0.4$, $0.5$, $0.6$, $0.8$, $1$, $2$, see Fig.~\ref{fig05}. As can be seen from this figure, the reported results for $c(T)$ and $\chi(T)$ are in fact robust with respect to the chosen number of thermalisation steps.

Figures \ref{fig03} and \ref{fig04} show the temperature dependence for different quantities as they follow from quantum Monte Carlo simulations to illustrate how the thermodynamic limit emerges as ${\cal N}$ grows. 
In these figures we also show the results for ${\cal N}=1$ obtained by exact diagonalization: They are in a perfect agreement with the ones obtained by quantum Monte Carlo simulations.

Monte Carlo simulations are usually accompanied by a finite-size scaling analysis. For instance, the ordering susceptibility $\chi$, which in the thermodynamic limit diverges according to a power law $\chi\propto\vert T-T_c\vert^{-\gamma}$, $T\to T_c$, exhibits finite-size scaling according to
\begin{eqnarray}
\label{27}
\chi(T)\approx 
{\cal L}^{\frac{\gamma}{\nu}}\tilde{\chi}\left(\left(T-T_{c}\right){\cal L}^{\frac{1}{\nu}}\right),
\end{eqnarray}	
see Refs.~\cite{Mueller-Krumbhaar1986,Binder1989}.
In the bottom panel of Fig.~\ref{fig04} we plot $T\chi(T){\cal L}^{-\gamma/\nu}$ against $\left(T-T_{c}\right){\cal L}^{1/\nu}$ to illustrate how data collapse in agreement with Eq.~(\ref{27}), in which we chose $T_c=0.334$, $\gamma=1.39$, $\nu=0.71$, see Ref.~\cite{Kivelson2024}. Besides, we show how a power-law divergence $\propto\vert \tau\vert^{-\gamma}$, $\tau=(T-T_c)/T_c\to \pm 0$, emerges as ${\cal L}$ increases; for that we also present the curve $500\vert\tau\vert^{-\gamma}$ (dash-dotted), see the inset in the bottom panel of Fig.~\ref{fig04}.

\section{Results and discussion}
\label{s4}

\begin{figure}
\includegraphics[width=0.995\columnwidth]{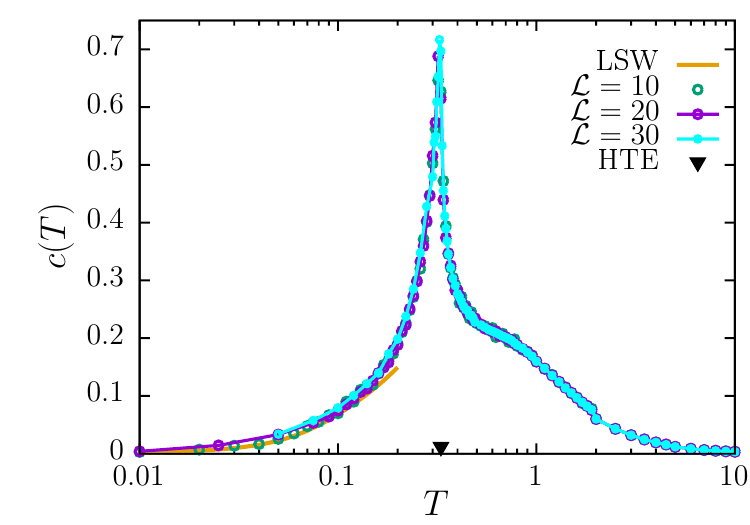}  
\includegraphics[width=0.995\columnwidth]{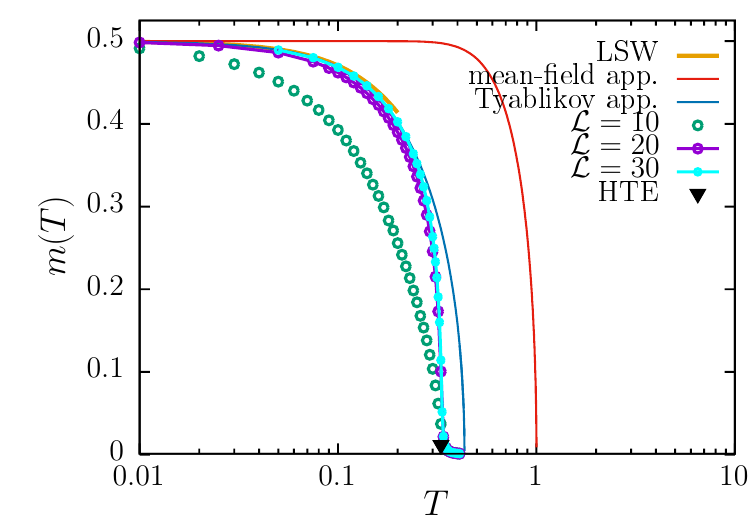}   
\caption{Thermodynamics of the hyperkagome-lattice $S=1/2$ Heisenberg ferromagnet: Temperature dependence of (top) specific heat and (bottom) magnetization.  
We report linear spin-wave results, Green's function calculations (mean-field and Tyablikov approximations), quantum Monte Carlo simulations for large enough systems with the symmetry-breaking field is $h=10^{-4}$, and high-temperature expansion series analysis for $T_c$.}
\label{fig06}
\end{figure}

Let us discuss the thermodynamic quantities for the hyperkagome-lattice $S=1/2$ Heisenberg antiferromagnet as they are predicted by different methods, see Fig.~\ref{fig06} and Table~\ref{tab1}.
Conventional linear spin-wave theory gives the standard power-law decay exponents for the internal energy and magnetization as $T\to 0$; the prefactors are related to the lattice geometry, which determines the coefficient for the (quadratic) spin-wave decay in the $\Gamma$ point, see Fig.~\ref{fig02}. The double-time temperature Green's function method within the mean-field and Tyablikov approximations yields qualitatively correct results not only at low temperatures, but also at intermediate temperatures. This method predicts a qualitatively reasonable temperature dependence of the magnetization in the ordered phase and the value of the Curie temperature: $T_c/\vert J\vert=1$ (mean-field approximation, Eq.~(\ref{18})) and $T_c/\vert J\vert\approx0.434$ (Tyablikov approximation, Eq.~(\ref{23})), see Table~\ref{tab1}. These results underestimate temperature fluctuations and the true Curie temperature is lower. Thus, high-temperature expansion series for the uniform susceptibility $\chi(\beta)$ up to $\beta^{16}$ \cite{Singh2012} leads to a lower value of  $T_c$: $T_c/\vert J\vert \approx 0.338\pm0.044$, whereas quantum Monte Carlo simulations suggest that $T_c/\vert J\vert \approx 0.330\pm0.004$ (according to the magnetization and the susceptibility, see Fig.~\ref{fig04}). High-temperature expansion series are valid not only in the disorder phase, but also can be used to determine the critical behavior around the Curie temperature. 
The approximate value of $\gamma$ exponent found from relatively short series \cite{Singh2012} in Sec.~\ref{s33}, $\gamma\approx 1.44$, is too crude for an unambiguous statement about the universality class, but it is close to the three-dimensional Heisenberg universality class \footnote{Critical exponents for the three-dimensional Heisenberg universality class rounded to two decimal places according to Ref.~\cite{Kivelson2024} are as follows: $\alpha=-0.12$, $\beta=0.36$, $\gamma=1.39$, $\delta=4.91$, $\eta=0.04$, and $\nu=0.71$.}.
Finally, the most informative findings come from the quantum Monte Carlo simulations for the system size up to $N=768\,000$ sites (${\cal N}=64\,000$ unit cells): Temperature dependence for various thermodynamic quantities refer to the whole temperature range, including the low-temperature ordered phase, the critical region around $T_c$, and the high-temperature disordered phase.
Interestingly, the specific heat $c(T)$ exhibits a sloped shoulder above $T_c$, see Figs.~\ref{fig03}, \ref{fig05}, \ref{fig06}. We are not aware of such a feature for other Heisenberg ferromagnets and this issue needs further study. However, it might be worth noting already here that since $c(T)$, but not $\chi(T)$, shows a shoulder as $T$ is about $0.4\ldots0.8$, one may associate this with some high-energy nonmagnetic excitations. For the antiferromagnetic sign of the exchange interaction they correspond to the low-energy states, which might be  responsible for intricate behavior of the hyperkagome-lattice $S=1/2$ Heisenberg antiferromagnet as $T$ tends to 0.

\begin{table}
\caption{\label{tab1} The Curie temperature $T_c$ as it follows from different approaches.}
\begin{ruledtabular}
\begin{tabular}{ll}
mean-field approximation          & $1$     \\
Tyablikov approximation           & $0.434$ \\   
high-temperature expansion series & $0.338\pm0.044$ \\  
quantum Monte Carlo simulations	  & $0.330\pm0.004$ \\
\end{tabular}
\end{ruledtabular}
\end{table}

As has been mentioned in Sec.~\ref{s1} and in the previous paragraph,
a model with the {\em ferromagnetic} sign of exchange interaction at finite temperatures has some indication 
of its antiferromagnetic counterpart at low temperatures. In particular, intricate low-temperature states of frustrated quantum spin system should show up at finite temperatures as the sign of exchange interaction changes.  
To illustrate such a correspondence, we compare the $S=1/2$ Heisenberg model on two three-dimensional lattices both with coordination number 4: The hyperkagome lattice and the diamond lattice. Antiferromagnetic Heisenberg exchange interaction for the former lattice is frustrated, whereas the latter bipartite lattice supports the N\'{e}el order. Now, the Curie temperature $T_c$ for the diamond-lattice $S=1/2$ Heisenberg ferromagnet is 
$0.447 \pm 0.001$ (high-temperature expansion) \cite{Oitmaa2018},
$\approx0.445$ (high-temperature expansion) \cite{Kuzmin2019},
or
$0.444\,47(5)$ (quantum Monte Carlo simulations) \cite{Baerwolf2025},
i.e., $T_c\approx 0.44$ is substantially higher than $T_c\approx 0.33$ for the hyperkagome-lattice case.

Finally, with our study, we have illustrated that the larger coordination number is, the better agreement with mean-field approach:
The mean-field $T_c$ for the simple-cubic and pyrochlore lattices (coordination number is 6) is $3/2$ that exceeds the quantum Monte Carlo results 0.839(1) \cite{Wessel2010} (simple-cubic) and 0.718 \cite{mueller2017} (pyrochlor) by 79\% and 109\%, respectively.
Now, the mean-field $T_c$ for the diamond and hyperkagome lattices (coordination number is 4) is $1$ that exceeds the quantum Monte Carlo results 0.44 (diamond) and 0.33 (hyperkagome) by about 125\% and 200\%, respectively.

\section{Summary}
\label{s5}

In the present paper, we have examined the thermodynamics of the hyperkagome-lattice $S=1/2$ Heisenberg ferromagnet which, to the best of our knowledge, have not been studied so far.
To investigate thermodynamic properties of this model we have used four complimentary approaches: The linear spin-wave theory, which gives thermodynamics at low temperatures, the Green's function method  augmented by the mean-field or Tyablikov approximation, which gives approximate thermodynamics up to the Curie temperature, high-temperature expansion series, which yield results only at high and intermediate temperatures, and quantum Monte Carlo simulations, which work for all temperatures but are always performed for finite systems (in our study, up to ${\cal N}=40^3$ unit cells).

In general, the studied quantum Heisenberg ferromagnet exhibits the properties which are qualitatively similar to such properties of other three-dimensional quantum Heisenberg ferromagnets (power-law exponents for thermodynamic quantities as $T\to 0$, a finite Curie temperature $T_c$, the critical behavior of the three-dimensional Heisenberg universality class around $T_c$, or paramagnetic behavior as $T\to\infty$). However, the peculiarity of the hyperkagome lattice (low coordination number, geometrical frustration, i.e., conflicting antiferromagnetic interactions) leads to some quantitative differences, e.g., the value of $T_c$ is rather small: $T_c\approx 0.33\vert J\vert$. Another unusual feature is a sloped shoulder in the $c(T)$ shape in the paramagnetic phase.

Although, the theoretical results presented in our paper should prove valuable in understanding the effect of lattice geometry on the observable properties of quantum  Heisenberg ferromagnet materials, there are to our knowledge, no experimental results yet available. However, there are several hyperkagome-lattice quantum Heisenberg antiferromagnets: Besides the already mentioned iridate Na$_4$Ir$_3$O$_8$, where Ir$^{4+}$ ions form the hyperkagome lattice, there are other  compounds, see, e.g., Refs.~\cite{Khatua2022,Shamoto2023,Yamauchi2024}. The sign of exchange interaction between magnetic ions arises from specific exchange mechanisms (superexchange, double exchange, etc. \cite{Fazekas1999}) and we may hope that with progress in materials science and synthesis of new compounds this lack of experimental data for the hyperkagome-lattice $S=1/2$ Heisenberg ferromagnet would be resolved in the future.

\section*{Data availability statement}

The data that support the findings of this study are available from the authors upon reasonable request.

\section*{Acknowledgements}

The authors are thankful to the Armed Forces of Ukraine for protection since 2014, and especially since February 24, 2022.
This project is funded by the National Research Foundation of Ukraine (2023.03/0063, Frustrated quantum magnets under various external conditions).
The authors thank Taras Hutak, Taras Verkholyak, and Petro Sapriianchuk for discussions and useful comments.
O.~D. is grateful to Joachim Stolze and G\"{o}tz Uhrig for hospitality at TU Dortmund University in November 2023.  

\bibliography{hyperkagome_refs_5}

\end{document}